\documentclass[a4paper,11pt]{article}%
\pdfoutput=1 

\usepackage{graphicx}
\usepackage{amsthm}
\usepackage{amsmath}
\usepackage{amsfonts}
\usepackage{amssymb}
\usepackage{color}
\RequirePackage[colorlinks,citecolor=blue,urlcolor=blue]{hyperref}

\usepackage{multirow}
\usepackage{anysize}
\marginsize{3.4cm}{3.4cm}{3.2cm}{3.2cm}

\renewcommand{\Pr}{\mathbb{P}}

\def\bft{\mathbf t}

\def\bfsigma{{\mbox{\boldmath${\sigma}$}}}

\newcommand{\E}{\operatorname{E}}

\numberwithin{equation}{section}

\begin{document}

\begin{center}

{ \Large \bf Estimating the prevalence of anemia rates among children under five in Peruvian districts with a small sample size}

\vspace{20pt}

{\bf Anna  Sikov}$\,^{b,c}$  and   {\bf Jos\'e Cerda-Hern\'andez}$\,^{a,c}$   

\vspace{20pt}

{\footnotesize


$^a$~Department of Engineering Economics, National Engineering University,\\
E-mail: jcerdah@uni.edu.pe\\
$^b$~Department of Engineering Statistics, National Engineering University\\
E-mail: asikov@uni.edu.pe\\
$^c$~Econometric Modelling and Data Science Research Group -- UNI\\
}

\vspace{30pt}

\end{center}

\begin{abstract}
In this paper we attempt to answer the following question: ``Is it possible to obtain reliable estimates for the prevalence of anemia rates in children under five years in the districts of Peru?''  Specifically, the interest of the present paper is to understand to which extent employing the basic and the spatial Fay-Herriot models can compensate for inadequate sample size in most of the sampled districts, and whether the way of choosing the spatial neighbors has an impact on the resulting inference. Furthermore, it is raised the question of how to choose an optimal way to define the neighbours. We present an illustrative analysis   using the data from the Demographic and Family Health Survey of the year 2019, and the National Census carried out in 2017. 	
\bigskip

\end{abstract}

\newpage

\section{Introduction}\label{sec1}

\begin{figure}[h!]
	\centering
	\includegraphics[scale=0.5]{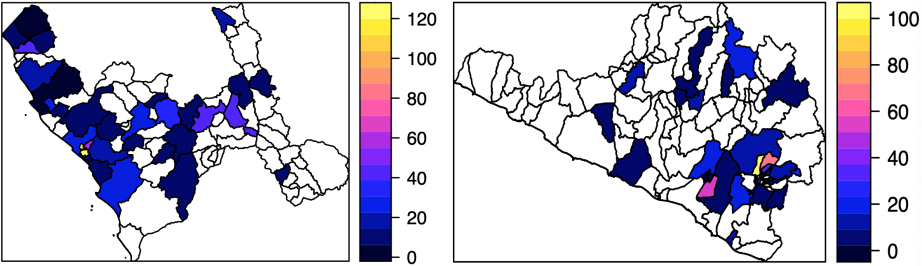}
	\caption{ENDES data: sample size in the districts of the dapartments of La Libertad (the left panel) and  Arequipa (the right panel), where the blank districts do not have available data.  }\label{muestra}
\end{figure}

The prevalence of anemia in young children is an important public health problem. According to the World Health Organization (WHO), anemia is a condition in which the number of red blood cells or the haemoglobin concentration within them is lower than normal, which can cause symptoms such as fatigue, weakness, dizziness and shortness of breath, among others (\cite{OMS2011}, \cite{WHO2004}). 
For this reason, reduction of prevalence of anemia is one of the priorities of the health policies of the Peruvian state. According to ``The National Plan for reduction and control of Maternal and Child Anemia and Chronic Child Malnutrition in Peru: 2017-2021'', presented by the  Ministery of Health, the target level was the reduction to 19\% of anemia in children by the end of 2021. Nonetheless, the prevalence of anemia, reported in 2018 was still 43.5\%, which corresponds to a reduction of 3.3\%, compared to the rates, observed in 2014 (\cite{MINSA2014}, \cite{MINSA2017}). Evidently, at the current rate of reduction the targeted level of 19\% will be attained only by the year 2050. In order to combat the problem of anemia in childhood, the Peruvian Government has implemented various social programs, such as ``Vaso de leche'', ``Juntos'' and ``Qali Warma'', the objetive of which is to reduce the prevalence of anemia and malnutrition in childhood. One of the most important aims of these programs is to quantify their impact on the reduction of the prevalence of anemia and malnutrition so as to optimize their costs and benefits (see \cite{Alcazar2012} for detailes). 
In order to evaluate this impact, good estimates of the percentage of anemic children are needed. However, in the case of Peru, obtaining these estimates, typically presents the most challenges, since there are many remote disticts, especially in mountainous regions, which are generally not included in the sample of the surveys due to logistic problems and limited budget; others have a very small sample size (see Figure \ref{muestra}). We will see below that a possible remedy to this problem would be to use spatial models, which exploit spatial correlations between the neighboring areas. However, populated areas in Peru are mostly located in mountainous regionsin, and therefore their location can be represented by three coordinates (longitude, latitude and altitude), in contrast to the proposed methods in the literature that use only the first two coordinates. Another problem is that application of the spatial Fay-Herriot model requires definition of the spatial neghbors which is completely subjective. In this study we address the question:  ``Is it possible to obtain reliable estimates for the prevalence of anemia rates in children under five years in the districts of Peru?''  in the presence of the above-mentioned problems. 

\medskip

In this article we utilize the two following sources of data: 1- the data provided by the Demographic and Health Survey- the ENDES, carried out by the National Institute of Statistics and Informatics in 2019 (\cite{INEI}) and 2- the data, obtained from the national census, carried out in 2017. The main objective of the national surveys like the ENDES is to describe some selected population characteristics such as health, employment and unemployment, education, household income and expenses, poverty etc. However one of the common problems of these surveys is that their corresponding sampling design is usually more appropriate for representing characteristics of the entire population, or of large subgrups, such as urban or rural population, the population of major geographical regiones, etc. Nonetheless, as noted by \cite{Rao2015}, more and more policy makers are demanding estimates for small domains to use them in the elaboration of policy decisions. In the case of the ENDES, inference at more disaggregated levels, such as provinces or districts is generally not reliable, since at these levels the areas may have small or null sample size. Namely, some of the areas of interest are usually not included in the sample, while the others do not have a sufficient number of observations in order to provide reliable direct estimates, based only on the area-specific sample data. As noted previously, in the case of Peru, the problem is even more pronounced due to limited logistics support and resources. For instance, in Puno region, only 34.5\% of the districts data regarding the prevalence of anemia is available. Furthermore, 65.8 \% of these districts have less than 10 observations. 
\medskip

In order to solve the problem of small sample sizes, the governmental entities like the statistical office of the European Union, United States Census Bureau among many others, utilize the basic Fay-Herriot model \cite{FH},  which is the area level model (distrit-level in our case). Based on this approach, the area level predictions are constructed as a linear combination of standard design-based estimates and indirect model-dependent estimates, where the corresponding regression model incorporates the auxiliary information, which is generally available from the census, administrative records or some other source of data, thus ``borrowing strength'' across other areas. Thereby, the basic Fay-Herriot model allows the areas to be linked through the vector of the regression coefficientes, compensating for their small sample sizes. The variation, which is not explained by the auxiliary variables, is accounted for by the corresponding area-specific random effects. In the case of the basic Fay-Herriot model, these effects are assumed to be independent. A limitation of the basic model is that it is not designed to handle the data that exhibit spatial dependence \cite{Moran} between the areas, which is the typical problem, arising in the data, collected from socio-economic surveys like the ENDES. In such situations, many authors (see for example, \cite{Cress1993}, \cite{MMM2013}, \cite{Petrucci}, \cite{Pratesi2008}, \linebreak \cite{Singh2005}) advocate the use of the natural extension of the basic model: the spatial Fay-Herriot model, which incorporates the information about geographical proximity of the areas which, in turn, is utilized to determine the covariance structure of the random effects of the spatially linked areas. More specifically, the  random effects are modelled by a simultaneously autoregressive model (SAR), which is characterized by a spatial autoregressive coefficient and a proximity matrix (see  \cite{Ansel1992}, \cite{Banerjee} and \cite{Cress1993} for more details). In this way, the expected value of a random effect of a specific area is defined as a linear combination of random effects of the neighboring areas. A drawback of this model is that it contains some degree of subjectivity, since it depends on the definition of the neighbours, which is aparently not unique. In addition, it should be noted,  that including spatial correlation into the model will not result in considerable gain in efficiency if this correlation is not substantially strong (\cite{Danny2002}). 
\medskip

In order to predict the area-specific characteristic of interest, Fay and Herriott (1979) develop the Best Linear Unbiased Predictor (BLUP). As mentioned above, this predictor constitutes a  composite estimator, which is derived as the weighted average of the direct area-specific estimator and a corresponding sintetic regression estimator. However, the BLUP can only be obtained if the variances of the random area-specific effects are known. In real applications, this is not always the case. If the variances are unknown, they are substituted by their corresponding estimates, obtained by maximum likelihood, restricted maximum likelihood or by a method of moments (\cite{FH}, \cite{Kackar}, \cite{Prasrao}, \cite{Rao2015}). The resulting predictor is the \textit{empirical} BLUP (EBLUP)(\cite{FH}). In the case of a spatial Fay-Herriot model, a Spatial Best Linear Unbiased Predictor (SBLUP) is used (see \linebreak \cite{Pratesi2008} for details). Replacing the unknown variance and autoregressive parameters by their corresponding estimates in the SBLUP leads to the \textit{empirical} SBLUP (SEBLUP).
\medskip

In this article we apply the basic and the spatial Fay-Herriot model in order to predict the percentage of anemic children under 5 years in the districts in Peru. Our main interest is to compare and to evaluate the performance of district-level predictors EBLUP and SEBLUP of the prevalence of anemia rates in the situation where the sampling design is inadequate in the sense that most districts are either not sampled or have a very small sample size, which is a typical problem in emerging and developing countries. As already mentioned, application of the spatial Fay-Herriot model is associated with some degree of subjectivity, introduced by definition of the neighbors. In order to address this issue we conduct a sensitivity analysis of the results to various definitions to the neighbours (see Section 4.4). This analysis is helpful to define the optimal choice of the neighbors.
Another complication that arises in our case is that each district has an additional dimension, namely the altitude. In Section 4 we consider how this additional coordinate can be aggregated in the definition of the neighboring districts. Next, we compute the mean square error for the aforementioned predictors. In the case of the basic Fay-Herriot model, we use the Prassad and Rao estimate  \cite{Prasrao} for the means square error, and in the case of the spatial Fay-Herriot we implement the parametric and non-parametric bootstrap, developed in \cite{Molina2009}. 
\medskip

The rest of the paper is organized as follows. In Section 2 the basic and the spatial Fay-Herriot models are presented. In Section 3 we briefly describe the problem of estimation of the MSE and provide some references to the most important works in this area. Section 4 illustrates a real data application. In this section the problem of subjectivity of the choice of neighboring areas, as well as the three-dimensional-coordinates problem are addressed. Finally, Section 5 provides some conclusions.

\section{Small Area Estimation Models}\label{sec2}

\subsection{Basic Fay-Herriot model}

Let  $Y_i$ denote the direct area-level estimate of the characteristic of interest in the $i-$th area, where $i,\; ~ i=1,...,D$ and $D$ is the total number of the areas with available data, and  $\theta_i$ donotes the corresponding true value of this characteristic. We suppose that $Y_i$ is design unbiased for $\theta_i$. Denote by $X_i=(x_{i1},...,x_{ip})$ the vector of $p$ auxiliary area-level covariates, which can usually be obtained from census or administrative sources. Then, the Fay-Herriot model is defined as follows
\begin{equation}\label{FH}
	Y_i=\theta_i+e_i; \quad \theta_i=X_i \beta+ u_i,
\end{equation}

\noindent Here $e_i \sim N(0,\sigma^2_i)$ are the errors of the direct estimates and $u_i \sim N(0, \sigma_u^2)$ are the area-level random effects, that represent the variability of the $\theta_i$'s  that is not explained by auxiliary variables, where $cov(e_i,e_j)=cov(u_i,u_j)=0$ if $i \ne j$ and $cov(e_i,u_j)=0 \quad \forall i,j$; $\beta$ is the vector of the coefficients that expresses the association between $\theta=(\theta_1,...,\theta_D)^t$ and $X=(X_1,...,X_D)^t$. It is assumed that the sampling error variances $\sigma^2_i$ are known. This assumption is customary, since the design variance of the sampling errors can usually be estimated from the observed data. Note that the coefficients $\beta$ do not depend on the  area. Specifically, the association between $X_i$ and $\theta_i$ is the same for all the areas, and hence the model-based estimate for the characteristic of interest in the $i$th area will incorporate the information about the other areas through the vector of coefficients $\beta$.   
\medskip

\noindent The model (\ref{FH}) can be rewritten as follows:
\begin{equation}\label{FH1}
	Y=X \beta+ u+e,
\end{equation}
\noindent where $Y=(Y_1,...,Y_D)^t$, $u=(u_1,...,u_D)^t \sim N(0,\Sigma_u)$, $e=(e_1,...,e_D)^t \sim N(0, \Sigma_e)$, such that  $\Sigma_u=\sigma^2_u I_{D}$  and $[\Sigma_e]_{ij}=\sigma^2_i I_{(i=j)},\quad i,j=1,...,D$. 

\noindent If $\sigma_u^2$ is known, $\theta_i, \quad i=1,...,D$ can be estimated using the Best Linear Unbiased Predictor (BLUP), developed in \cite{FH}, as follows.
\begin{equation}\label{BLUP}
	\hat{\theta}_i^{BLUP}\left(\sigma_u^2\right)=X_i \hat\beta \left(\sigma_u^2\right)+ \hat u_i \left(\sigma_u^2\right),
\end{equation}

\noindent Here, 

\begin{equation}\label{beta}
	\hat \beta \left(\sigma_u^2\right)=\left(X^t \left[{V}(\sigma_u^2)\right]^{-1}X \right)^{-1}X^t \left[{V}(\sigma_u^2)\right]^{-1}Y,
\end{equation}

\begin{equation}\label{UFH}
	\hat u_i \left(\sigma_u^2\right)=E\left(u_i \mid Y_i\right) = \gamma_i\left(\sigma_u^2\right)\left(Y_i-X_i \hat\beta \left(\sigma_u^2\right) \right)
\end{equation}
where 

$V\left(\sigma_u^2\right)=Var(u+e)=\Sigma_u+\Sigma_e$  and 

$\gamma_i \left(\sigma_u^2\right)=\displaystyle\frac{\sigma_u^2}{\sigma_i^2+\sigma_u^2}$.  
\medskip

\noindent Alternatively, the predictor (\ref{BLUP}) can be presented as
\begin{equation}\label{BLUP1}
	\hat{\theta}_i^{BLUP}\left(\sigma_u^2\right)=\gamma_i \left(\sigma_u^2\right) Y_i + \left(1-\gamma_i \left(\sigma_u^2\right)\right)X_i \hat\beta \left(\sigma_u^2\right)
\end{equation}

\noindent Note that the predictor (\ref{BLUP1}) constitutes a convex combination of the direct estimate $Y_i$ and the model-based estimate  $X_i\hat\beta$. Clearly, if the $i$th area does not have available data, its corresponding value of $\gamma_i$ is equal to zero, and therefore the prediction of $\theta_i$ for this area is equal to the model-based estimator.   
\medskip

In most real data applications, the value of the parameter $\sigma_u^2$ is unknown. In this case, $\sigma_u^2$ can be estimated by means of maximum likelihood (ML), restricted maximum likelihood (REML), the method of moments, developed by Prasad and Rao (1990) for the Fay-Herriot model (see \cite{Prasrao}), or the method, proposed by Fay and Herriot (see \cite{FH} for details).
\medskip

\noindent The log-likelihood function is obtained as
\begin{equation}\label{ML}
	l_{ML}(\beta,\sigma_u^2)=c-\displaystyle\frac{1}{2} \log \mid V \mid-\displaystyle\frac{1}{2} (Y-X\beta)V^{-1}(Y-X\beta)^t
\end{equation}
where $c$ is some constant and $V=V(\sigma_u)$. Given function is maximized with respect to $\sigma_u^2$, whereas the parameters $\beta$ are estimated as (\ref{beta}).
\medskip

\noindent The restricted log-likelihood function is defined as
\begin{equation}\label{REML}
	l_{REML}(\sigma_u^2)=c'-\displaystyle\frac{1}{2} \log\mid V \mid	-\displaystyle\frac{1}{2} \log \mid X^t V^{-1}X \mid-\displaystyle\frac{1}{2}Y^tPY, 
\end{equation}
where $c'$ is some constant, $V=V(\sigma_u)$ and $P=V^{-1}-V^{-1}X(X^t V^{-1}X)^{-1}X^tV^{-1}$.
\medskip

\noindent Contrary to the ML, the REML takes into account the loss of degrees of freedom due to estimation of the parameters $\beta$, and consequently, it is advantageous in the case of small sample sizes (\cite{Molina2009}, \cite{Rao2003}, \cite{Rao2015}).  
\medskip 

\noindent The method of moments estimate for $\sigma_u^2$ can be obtained as
\begin{equation}\label{Var_mom}
	\tilde\sigma_u^2=\displaystyle\frac{1}{D-p}\sum_{i=1}^{D} \left[ \left(Y_i-X_i\hat\beta_{OLS}\right)^2-\sigma_i^2(1-h_i) \right],
\end{equation}
where $\hat\beta_{OLS}=(X^tX)^{-1}X^t Y$, $h_i=X_i (X^tX)^{-1}X_i^t$ and $p$ is the number of auxiliary area level covariates in the model (\ref{FH}).
However, since the value of $\tilde\sigma_u^2$ can take a negative value, the estimate for $\sigma_u^2$ is given by
\begin{equation}\label{Var_mom1}
	\hat\sigma_u^2=\max\{  0, \tilde\sigma_u^2 \}.
\end{equation}
The estimate, proposed by Fay and Herriot (1979) (see  \cite{FH}) is derived as an iterative solution of the equation 
\begin{equation}\label{Var_FH}
	\sum_{i=1}^{D} \displaystyle\frac{(Y_i-X_i\beta^*)^2}{\sigma_u^{*2}+\sigma_i^2}=(D-p)
\end{equation}
where $\beta^*$ is obtained from (\ref{beta}).
\medskip

\noindent It is important to emphasize that all mentioned estimates for $\sigma_u^2$ are translation invariant, that is, have the following properties (see \cite{Kackar} for more details):

\begin{enumerate}
	\item $\hat\sigma_u^2(Y)=\hat\sigma_u^2(-Y)$
	\item $\hat\sigma_u^2(Y-Xa)=\hat\sigma_u^2(Y)$, $\forall \; a \in R^p$  and  $\forall \; Y$.
\end{enumerate}

\noindent Kackar and Harville (1984) \cite{Kackar} show that the empirical BLUP  $\hat{\theta}_i^{EBLUP}$, which is defined in \cite{FH} as 

\begin{equation}\label{EBLUP}
	\hat{\theta}_i^{EBLUP} \left(\hat\sigma_u^2\right) = \gamma_i \left(\hat\sigma_u^2\right) Y_i+\left(1-\gamma_i\left(\hat\sigma_u^2\right) \right)X_i \hat\beta \left(\hat\sigma_u^2\right),
\end{equation}
\medskip 

\noindent is unbiased for $\theta_i$ if a consistent estimate $\hat\sigma_u^2$ is translate invariant. 
\medskip

\noindent As discussed previously, if the data present strong spatial correlations, a spatial Fay-Herriot model is a natural way to proceed. This model is described in the following subsection.

\subsection{Spatial Fay-Herriot Model}

The spatial Fay-Herriot model is defined as follows (see \cite{Pratesi2008} for more details):
\begin{equation}\label{EFH}
	Y=X\beta+u + e; \quad u=\rho W u+ \epsilon,
\end{equation}
where $\epsilon=(\epsilon_1 ,...,\epsilon_D)^t \sim N(0,\Sigma_\epsilon)$ such that $\Sigma_\epsilon=\sigma_\epsilon^2 I$, $\rho$ is the spatial autoregressive  coefficient (see \cite{Banerjee}, \cite{Cress1989} and \cite{Cress1993}), and $W$ is a matrix of non-negative spatial weights, the elements $w_{ij}$ of which define the spatial measure of proximity between the areas $i$ and $j$, such that $\forall \;  i=1,...,D$, $w_{ii}=0$ and $\sum_{j=1}^D w_{ij}=1$. As noted above, the weights $w_{ij}$ can be defined in a variety of ways. Typically, $w_{ij}$ depend on the definition of the neighbouring areas. However, it must be noted that, it is hard to formulate specific criteria to choose the "best" definition. Here we present a few common approaches to define neighboring areas of a specific area $i$ (the interested readers can refer to \cite{Ansel1992} and \cite{Cress1993} for more details). 

\medskip 
\begin{enumerate}
	\item Those areas, whose distance between their corresponding centroids and the centroid of the area of interest is within $L$ miles. For example, \linebreak \cite{Cress1989} define two areas as neighbours if the distance between their centroids is within 30 miles. 
	\item The $k$ nearest areas to the area of interest. 
	\item Areas that share a common boundary with the area of interest.
\end{enumerate}
\medskip

\noindent Clearly, it is important to use caution when defining the neighbors, since different definitions may produce different results.
\medskip


\noindent Now, the model (\ref{EFH}) can be written as: 
\begin{equation}\label{EFH1}
	Y=X \beta+ (I-\rho W)^{-1}\epsilon + e= X \beta+\nu,\quad \nu \sim N(0,G), 
\end{equation} 
where
$$G=\sigma^2_\epsilon \left[ (I-\rho W)^t (I-\rho W)  \right]^{-1} +\Sigma_{e}=\Omega+\Sigma_{e}.$$
Note that the matrix $G$ exists only if $(I-\rho W)$ is non-singular. 
\medskip

\noindent  Next, let $\phi=(\sigma_\epsilon^2 ,\rho)$ index the unknown model parameters, and \linebreak $b_i=(0,...,0,1,0,...,0)^t$ be a D-dimensional vector with value 1 in the $i$th position and 0 in all other positions. Therefore, the spatial BLUP (SBLUP) for $\theta_i$, is obtained as:
\begin{equation}\label{SBLUP}
	{\hat\theta}_i^{SBLUP}(\phi)=X_i \hat\beta(\phi)+ \hat u_i(\phi),
\end{equation} 
where
\begin{equation}\label{beta.esp}
	\hat\beta(\phi)=\left(X^t [{G}(\phi)]^{-1}X\right)^{-1}X^t [{G}(\phi)]^{-1}Y
\end{equation} 
and
\begin{equation}\label{u.esp}
	\hat u_i (\phi)=b_i^t\Omega^t(\phi) \left[{G}(\phi)\right]^{-1}\left(Y-X \hat\beta(\phi)\right)
\end{equation} 
The estimates of the unknown parameters $\phi$ can be obtained using ML or REML, where the covariance matrix $V$  in (\ref{ML}) or (\ref{REML}) is replaced by the matrix $G(\phi)$. Molina, Salvati and Pratesi (2009) \cite{Molina2009} warn about possible numeric problems, associated with optimization of the functions (\ref{ML}) and (\ref{REML}) in this case.  
\medskip

\noindent Replacing the parameters $\phi$ with there corresponding estimates, $\hat\phi$ in (\ref{beta.esp}) and in (\ref{u.esp}), we obtain the empirical SBLUP (SEBLUP) for $\theta_i$, which is given by
\begin{equation}\label{SEBLUP}
	{\hat\theta}_i^{SEBLUP}(\hat\phi)=X_i \hat\beta(\hat\phi)+ \hat u_i(\hat\phi),
\end{equation} 
The estimate (\ref{SEBLUP}) is unbiased for $\theta_i$ if $\hat\sigma_{\epsilon}^2$  and $\hat\rho$ are derived using ML or REML (see \cite{Kackar} for more details). 

\section{Estimation of the Mean Square Error of EBLUP and SEBLUP}\label{sec3}
In real applications, a natural question of interest is how to estimate the mean square error (MSE) of the predictors (\ref{EBLUP}) and (\ref{SEBLUP}). In this section we present a brief review of the main estimation methods that have been proposed in the literature to address this problem. We start with analizing the MSE of the BLUP (\ref{BLUP}). It can be easily shown that
\begin{equation}\label{ECM_BLUP}
	\begin{array}{ccl}
		MSE\left({\hat\theta}_i^{BLUP}(\sigma_u^2)\right) \!\! & =& \!\! \gamma_i(\sigma_u^2)\sigma_i^2+\left(1-\gamma_i(\sigma_u^2)\right)^2 X_i Var \left(\hat\beta(\sigma_u^2)\right)X_i^t \vspace{0.2cm} \\ 
		\!\! & = & \!\!  g_{1i}(\sigma_u^2)+g_{2i}(\sigma_u^2),
	\end{array}
\end{equation} 
where $X_i$ is the $i$th line of the matrix $X$ and $\hat\beta(\sigma_u^2)$ is the estimate for $\beta$, defined in (\ref{beta}). Note that, the component $g_{1i}(\sigma_u^2)$ corresponds to the sampling error, whereas $g_{2i}(\sigma_u^2)$ expresses the error associated with estimation of the parameters $\beta$. It is important to emphasize that
$g_{1i}(\sigma_u^2)=O(1)$
and
$g_{2i}(\sigma_u^2)=O\left(\displaystyle\frac{1}{D}\right)$
and therefore if the total number of areas $D$ is large, $MSE({\hat\theta}_i^{BLUP}(\sigma_u^2))\approx g_{1i}(\sigma_u^2)$. 
Obviously, $g_{1i}(\sigma_u^2)$ is smaller than $\sigma_i^2$, which is the MSE of the direct estimate.
In fact, $g_{1i}(\sigma_u^2)$ is substantially smaller than $\sigma_i^2$ if the value of $\sigma_u^2$ is small which occurs when good covariate information if available. The estimate for the MSE defined in (\ref{ECM_BLUP}) is obtained by replacing $\sigma_u^2$ with the estimate $\hat\sigma_u^2$, as follows.
\begin{equation}\label{EECM_BLUP}
	\mbox{mse}({\hat\theta}_i^{BLUP}(\sigma_u^2))=g_{1i}(\hat\sigma_u^2)+g_{2i}(\hat\sigma_u^2),
\end{equation} 

\noindent It should be noticed that (\ref{ECM_BLUP}) and (\ref{EECM_BLUP}) do not account for the error associated with the estimation of the parameter $\sigma_u^2$. 
It can be demonstrated that (see  \linebreak  \cite{Kackar} and  \cite{Harville1992}) if the sampling errors and the area-level random effects have a normal distribution, and the estimate for $\sigma_u^2$ is translation invariant, the MSE can be decomposed as:
\begin{equation}\label{ECM_EBLUP}
	MSE \! \left( \! {\hat\theta}_i^{EBLUP} \! (\hat\sigma_u^2) \! \right) \! = \! MSE \! \left( \! {\hat\theta}_i^{BLUP} \! (\sigma_u^2) \! \right) \! + \! E \! \left( \! {\hat\theta}_i^{EBLUP} \! (\hat\sigma_u^2) \! - \! {\hat\theta}_i^{BLUP} \! (\sigma_u^2) \! \right)^2
\end{equation} 
The second term in the expression (\ref{ECM_EBLUP}) represents the additional error which is the result of the estimation of the parameter $\sigma_u^2$. Contrary to the first term, the second term can not be expressed analitycally, and therefore, can only be obtained by approximation. 
If $\sigma_u^2$ is estimated by the method of moments, defined in (\ref{Var_mom}) and (\ref{Var_mom1}), the MSE of $\theta_i^{EBLUP}$ can be approximated utilizing the method proposed by Prasad and Rao \cite{Prasrao}, as follows:
\begin{equation}\label{ECM_PR}
	MSE({\hat\theta}_i^{EBLUP}(\hat\sigma_u^2)) \approx g_{1i}(\sigma_u^2)+g_{2i}(\sigma_u^2)+Var(\hat\sigma_u^2)g_{3i}(\sigma_u^2),
\end{equation} 
where $g_{3i}(\sigma_u^2)=\displaystyle\frac{(\sigma_i^2)^2}{(\sigma_i^2+\sigma_u^2)^3}$ 
and $Var(\hat\sigma_u^2)\approx \displaystyle\frac{1}{2D^2}\sum_{i=1}^{D} \left(\sigma_i^2+\sigma_u^2\right)^2$. 
\medskip

\noindent The authors demonstrate that in this case the estimate for the MSE can be obtained as
\begin{equation}\label{EECM_PR}
	\mbox{mse}\left({\hat\theta}_i^{EBLUP}(\hat\sigma_u^2)\right) = g_{1i}(\hat\sigma_u^2)+g_{2i}(\hat\sigma_u^2)+2\hat {Var}(\hat\sigma_u^2)g_{3i}(\hat\sigma_u^2),
\end{equation} 
and that the proposed estimate has the bias of order $o\left(\displaystyle\frac{1}{D}\right)$.\\
In \cite{Datta2005} the authors develop the estimate for the MSE of $\hat\theta_i^{EBLUP}$ in the case where $\sigma_u^2$ is estimated by (\ref{Var_FH}), as follows
\begin{equation}\label{EECM_Datta}
	\mbox{mse}\left({\hat\theta}_i^{EBLUP}(\hat\sigma_u^2)\right) = g_{1i}(\hat\sigma_u^2)+g_{2i}(\hat\sigma_u^2)+2\hat {Var}(\hat\sigma_u^2)g_{3i}(\hat\sigma_u^2)-g_{4i}(\hat\sigma_u^2),
\end{equation} 
where 

$$
\begin{array}{ccl}
	g_{4i}(\hat\sigma_u^2) &=& 2(1-\gamma_i (\hat\sigma_u^2))^2 \times \left[ D\sum_{i=1}^{D}\displaystyle\frac{1}{(\sigma_i^2+\hat\sigma_u^2)^2}-
	\left(\sum_{i=1}^{D}\displaystyle\frac{1}{(\sigma_i^2+\hat\sigma_u^2)}\right)^2 \right] \times \vspace{0.2cm} \\
	& &  \left( \sum_{i=1}^{D}\displaystyle\frac{1}{(\sigma_i^2+\hat\sigma_u^2)} \right)^{-3}
\end{array}
$$ 
The order of the bias of the estimate (\ref{EECM_Datta}) is $o(\displaystyle\frac{1}{D})$. 
\medskip

\noindent If $\hat\sigma_u^2$ is obtained using the method of ML or REML, the MSE of $\hat\theta_i^{EBLUP}$ can be estimated utilizing the approximation developed in \cite{Datta2000}. As in the previous cases the order of the bias of the proposed estimate is $o\left(\displaystyle\frac{1}{D}\right)$. 
\medskip

\noindent Alternatively, the MSE can be estimated with the same order of the bias utilizing resampling methods, such as the bootstrap and jackknife (see \cite{Chen2003}, \cite{Hall2006} and \cite{Jiang2002} among many others). 
\medskip

\noindent If the spatial Fay-Herriot model is used, an additional parameter $\rho$ is to be estimated. 
As noted previously, unknown parameters $\phi=(\sigma_u^2, \rho)$ can be estimated using the method of ML or REML. As in the previous case the MSE of $\theta_i^{SEBLUP}$ can be decomposed as ( \cite{Molina2009},  \cite{Pratesi2008} and \cite{Singh2005}):
\begin{equation}\label{ECM_SEBLUP}
	\begin{array}{ccl}
		\mbox{MSE} \! \left( \! {\hat\theta}_i^{SEBLUP}(\hat\phi) \! \right) \!\!\! &=& \!\!\! \mbox{MSE} \! \left( \! {\hat\theta}_i^{SBLUP} \! (\phi) \! \right) \! + \!
		E \! \left( \! {\hat\theta}_i^{SEBLUP} \! (\hat\phi) \! - \! {\hat\theta}_i^{SBLUP} \! (\phi) \! \right)^2 	\vspace{0.2cm} \\
		\!\!\! &=& \!\!\!
		g_{1i}(\phi)+g_{2i}(\phi)+g_{3i}(\phi),
	\end{array}
\end{equation} 
where the term $g_{1i}(\phi)$ represents the error produced by the estimation of the random effects and has the order $O(1)$, and the term $g_{2i}(\phi)$ represents the error produced by the estimation of the parameters $\beta$ and it is of the order $O \left(\displaystyle\frac{1}{D} \right)$ (see \cite{Singh2005}). 
If the parameters $\phi$ are estimated by means of REML, the estimate for the MSE is approximately unbiased and is given by

\begin{equation}\label{ecm_SEBLUP}
	mse({\hat\theta}_i^{SEBLUP}(\hat\phi)) \approx g_{1i}(\hat\phi)+g_{2i}(\hat\phi)+2g_{3i}(\hat\phi) 
\end{equation} 

\noindent If ML is used for estimation of $\phi$, the expression for the estimate of the MSE includes an extra term, which corrects for the additional bias of $g_{1i}(\hat\phi)$ (see  \linebreak  \cite{Molina2009}, \cite{Pratesi2008}, \linebreak \cite{Pratesi2009}, \cite{Singh2005} for details).
\medskip

The expressions of $g_{1i}(\phi)$ and $g_{2i}(\phi)$ can be obtained analytically  (computational details can be found in \cite{Singh2005}), whereas for the term $g_{3i}(\phi)$ which represents the error due to estimating the parameters $\phi$, no analytic form can be derived. In \cite{Pratesi2009} the authors propose a heuristic aproximation for $g_{3i}(\phi)$. Alternatively, a bootstrap method can be adopted in order to estimate $g_{3i}(\phi)$. Here, we present the parametric bootstrap, proposed by \cite{Molina2009}.

\begin{enumerate}
	\item Fit model (\ref{EFH}) to the original data $Y=(Y_1,...,Y_D)^t$ in order to obtain the estimates $\hat\phi=(\hat\sigma^{2}_u, \hat\rho)$  and  $\hat\beta$.
	\item Generate $B$ bootstrap samples, utilizing the model (\ref{EFH}) with the parameters estimated  in step 1, as follows. 
	\begin{enumerate}	
		\item Generate a vector $Z_1^b=(Z_{11}^b,Z_{12}^b,...,Z_{1D}^b)^t$ of independent variables, such that $Z_{1j}^b \sim N(0,1)$, $j=1,...,D$, $b=1,...,B$, and compute $\tilde{u}^b=\hat\sigma_u Z_1^b$, $u^b=(I-\hat\rho W)^{-1}\tilde u^b$.
		\item Generate a vector $Z_2^b=(Z_{21}^b,Z_{22}^b,...,Z_{2D}^b)^t$ of independent variables, such that $Z_{2j}^b \sim N(0,1)$, $j=1,...,D$,$b=1,...,B$ and compute $e^b=(e_1^b,...,e_D^b)$, where $e_j^b=\sigma_jZ_{2j}^b$. 
		\item Compute bootstrap area level characteristics of interest,  $\theta^b=X\hat\beta+u^b$, and bootstrap data $Y^{b}=\theta^b+e^b$.
	\end{enumerate}
	\item For each bootstrap sample, $Y^b$, $b=1,...,B$, reestimate $\phi$ and $\beta$, obtaining $\hat\phi^b$ and $\hat\beta^b(\hat\phi)$, where $\hat\phi^b$ is derived by application of ML or REML and the estimates $\hat\beta^b(\hat\phi)$ and $\hat\beta^b(\hat\phi^b)$ are computed using (\ref{beta.esp}), where $\phi$ is replaced by $\hat\phi$ and $\hat\phi^b$ respectively. 
	\item For each bootstrap sample, $Y^b$, $b=1,...,B$, compute $\hat\theta^{SBLUP,b}(\hat\phi)$ and \linebreak $\hat\theta^{SEBLUP,b}(\hat\phi^{b})$ as:
	$$\hat\theta^{SBLUP,b}(\hat\phi)=X\hat\beta^b(\hat\phi)+\Omega^t(\hat\phi)[G(\hat\phi)]^{-1}(Y^b-X\hat\beta^b(\hat\phi))$$
	and
	$$\hat\theta^{SEBLUP,b}(\hat\phi^b)=X\hat\beta^b(\hat\phi^b)+\Omega^t(\hat\phi^b)[G(\hat\phi^b)]^{-1}(Y^b-X\hat\beta^b(\hat\phi^b))$$  
	\item Now, the bootstrap estimate for $g_{3i}(\phi)$ is given by
	$$g_{3i}^{PB}(\hat\phi)=\displaystyle\frac{1}{B} \sum_{b=1}^B \left[\hat\theta_i^{SEBLUP,b}(\hat\phi^b)-\hat\theta_i^{SBLUP,b}(\hat\phi) \right]^2$$
\end{enumerate}
Another estimate for the MSE of the SEBLUP (\ref{SEBLUP}) was developed in  \linebreak \cite{Danny2005} and it is computed as 
\begin{equation}\label{ecmBoot}
	\begin{array}{ccl}
		\mbox{mse}(\theta_i^{SEBLUP}(\hat\phi)) &=& 2(g_{1i}(\hat\phi)+g_{2i}(\hat\phi)) \vspace{0.2cm} \\ 
		& & - 
		\displaystyle\frac{1}{B}\sum_{b=1}^B(g_{1i}(\hat\phi^b)+g_{2i}(\hat\phi^b))+g_{3i}^{PB}(\hat\phi)
	\end{array}
\end{equation}

\noindent Analogously, one can use a non-parametric bootstrap, developed in  \linebreak \cite{Molina2009}. In this case, the bootstrap random effects and the sampling errors are drawn from the empirical distribution of the predicted random effects and from the model residuals,  respectively. As noted by the authors, this method avoids the need of distributional assumptions and therefore, it is expected to be more robust to non-normality of any of the random components of the model.  

\section{A Case Study}
\subsection{Objectives of the study}
In this section we illustrate and study the performance of the basic and spatial Fay-Herriot models  using data collected as part of the Demographic and Health Survey- ENDES, carried out 
by the National Institute of Statistics and Informatics in 2019. The survey collects information on 
the topics such as anemia, nutrition, education, domestic violence among many others. The sampling units in this survey are households, which were sampled by a two-stage sampling design: at the first stage, a sample of localities was selected; at the second stage, a sample of dwellings was chosen within each of the selected localities. A household is defined as a group of people living in the same dwelling and sharing the same budget for food expenditure. In this study we focus on modeling the prevalence of anemia rates in children under five, per district. As it has been pointed out previously, these estimates are unreliable for most of the sampled districts. Our main aim is to study gain in precision of the estimates obtained by employing the aforementioned models. Specifically, we focus on the following two points. First, we address the question of choosing the neighbor criterion to be used. Second, we compare the MSE and the coefficient of variation of the predictors EBLUP and SEBLUP obtained by application of the basic and the spatial Fay-Herriot model, respectively. The auxiliary covariates used in the model are the characteristics of the district, obtained from the National Census carried out in 2017, as displayed in the following table.    

\begin{table}[h!]
	\caption{Description of the auxiliary variables} \label{tab:var}
	\centering
	\begin{tabular}{p{1.5cm} p{9.5cm}}
		\hline
		\textbf{Variable} & \textbf{Description of the variable} \\
		\hline
		Altitude  & The altitude of the district (height above sea level)\\
		Water & \% of dwellings with access to centralized water supply\\
		Water-days & \% of dwellings with access to potable water only several days per week\\
		Floor & \% of dwellings that have non-dirt flooring\\
		Internet & \% of dwellings with access to internet\\
		SIS & \% of the population that is affiliated with the Comprehensive Health Insurance (SIS)\\
		Uninsur. & \% of the population that do not have health insurance \\ 
		Refrig. & \% of households that have a refrigerator\\  
		Spanish & \% of native Spanish speakers\\ 
		Rural & \% of rural dwellings\\
		\hline
	\end{tabular}
\end{table}

Application of the basic (\ref{FH}) and the spatial (\ref{EFH1}) Fay-Herriot models to all the districts with available direct estimates resulted in a very poor fit. In order to remedy this problem, we divided all the districts into the following three groups: 1- the districts, where less than 30\% of the population live in poverty (a total of 585 districts, 281 sampled districts), 2- the districts where 30\%-55\% of the population live in poverty (a total of 671 districts, 297 sampled districts) and 3- the districts where more than 55\% of the population live in poverty (a total of 618 districts, 234 sampled districts), and fit the aforementioned models in each of the specified groups separately.
Next, we compare the estimators for the MSE of the EBLUP and SEBLUP, defined by (\ref{EBLUP}) and (\ref{SEBLUP}) correspondingly. In the case of the EBLUP, we utilize the estimator proposed by \cite{Prasrao}, defined in (\ref{ECM_PR}). In order to obtain the estimator for the MSE of the SEBLUP we use the parametric and non-parametric bootstrap, proposed in \cite{Molina2009}. 

\subsection{Definition of the neighboring districts} 
In what follows the neighbors of a specific district are defined in two steps. In the first step, $K_1$ nearest neighbors are chosen, using districts' latitude and longitude, where $K_1=3,4,...,10$. 
It should be noticed that another two ways to define the neighbors, mentioned in section 2.2 are inapplicable in our case due to a large number of nonsampled districts.
In the second step we use the difference in altitude as the measure of proximity between each of the $K_1$ previously selected districts and the district of interest. In this step we choose $K_2 \leq K_1$ "closest" districts. The spatial weights of each of the $K_2$ districts selected in the second step is equal to $1/K_2$, while the spatial weights of all other district are equal to 0. For this study we use $K_2=1,...,K_1$. Then, for each pair $(K_1,K_2)$ we analize the fit of the spatial Fay-Herriot model. Specifically, we study the behavior of the estimator for the variance of the model errors, $\hat\sigma^2_\varepsilon$ as a function of $(K_1,K_2)$. Obviously, the optimal definition of the neighbor corresponds to the values of $K_1$ and $K_2$ which results in the smallest value of $\hat\sigma^2_\varepsilon$. In should be noted that in the second step the proximity (or similarity) between the neighboring districts can be expressed using other variables, for example, the poverty level or human development index in the district. This additional information can be potentially useful, especially in the case where many areas have small o very small sample size. 
In this study in addition to the variable ``Altitude" we use the variables ``Poverty" and ``Extreme Poverty" which stand for the percentage of the population living in poverty and extreme poverty, respectively. 

\subsection{Fitting the basic Fay-Herriot models}

Initially, we present the results of fitting the basic Fay-Herriot models. Table \ref{tabl2} shows the estimated coefficients $\hat\beta$ of the model and their corresponding $p$-values, as obtained when fitting the  model separately to each of the three defined groups of the districts.

\begin{table}[h]
	\caption{Estimators of the coefficients $\beta$ of the basic Fay-Herriot model}\label{tabl2}
		\begin{tabular*}{\textwidth}{@{\extracolsep{\fill}}lcccccc@{\extracolsep{\fill}}}
			\hline
			& \multicolumn{2}{@{}c@{}}{Less than 30\%} & \multicolumn{2}{@{}c@{}}{30\%-55\%} & \multicolumn{2}{@{}c@{}}{More than 55\%}\\  
			& Estimator & p.value & Estimator & p.value & Estimator & p.value \\
			\hline 
			Water  & --- & --- & -0.15557 & 0.0002 & -0.10699 & 0.0287 \\
			Water-days  & 0.08827 & 0.0480  & --- & --- & --- & --- \\
			Floor  & -0.13376 & 0.0023 & --- & --- & --- & ---\\
			Refrig.  & --- & --- & -0.33600 & $<0.0001$ & -0.18277 & 0.0015\\
			Internet  & -0.39027 & $<0.0001$ & --- & --- & --- & ---\\
			Spanish  & -0.25204 & $<0.0001$ & -0.17888 & $<0.0001$ & -0.23563 & $<0.0001$\\
			SIS  & --- & --- & -0.18834 & 0.0101 & --- & ---\\
			Uninsur. & --- & ---  & ---  & --- & 0.45826 & 0.0001\\
			Altitude & 0.00002 & 0.0027  & --- & --- & 0.00002 & 0.0291\\
			Rural & --- & --- & -0.07049 & 0.0484 & --- & ---\\
			\hline
		\end{tabular*}
\end{table}

Table \ref{tabl2} indicates that the prevalence of anemia in a district is apparently associated with the variables that reflect the poverty level of that district. It should be noted that many other auxiliary variables that also reflect the poverty level in a district, such as the percentage of dwellings with concrete walls, the percentage of dwellings with access to centralized hygiene system, the percentage of illiterate population etc., were initially included in the model, however their corresponding coefficients were not significant. 

\subsection{Sensitivity Analysis}
In this Section we conduct a sensitivity analysis to investigate the impact of selecting the neighboring districts. To this end, the spatial Fay-Herriot model was fitted with $K_1=1,...,10$ and $K_2=1,...,K_1$ neighbors, as explained in Section 4.2. The figures in Tables \ref{tabl3}-\ref{tabl5}  suggest that the results are sensitive to the way in which the neighbours were defined. Furthermore, it should be noted that the estimators of the parameter $\rho$ vary quite widely with the choice of $K_1$ and $K_2$ (from 0.15 to 0.87). These results demonstrate that the way of choosing of the neighbors can dramatically alter inferences. In this situation we recommend using the values of $K_1$ and $K_2$ that correspond to the minimal value of $\hat\sigma_{\epsilon}^2$.  
The results displayed in the tables, illustrate that the optimal choice of the neighbors in the case of the districts of the first two groups is $K_1=3$ and $K_2=2$, whereas for the third group  the optimal values are $K_1=7$ and $K_2=3$. At the same time, the tables show that if the variable ``Altitude" is not utilized, which implies $K_2=K_1$, the optimal value of $K_1$ in the case of the first two groups is $K_1=2$, while for the third group $K_1=7$. Comparing the corresponding magnitudes of $\hat\sigma_{\epsilon}^2$, it can be observed that incorporating the variable ``Altitude" leads to a minor reduction of 5\% (from 0.0041 to 0.0039 and from 0.0042 to 0.0040) in the first and the third group, and of 22\% (from 0.0027 to 0.0022) in the second group.

\begin{table}[h!]
	\centering
	\caption{The values of $\hat\sigma_{\epsilon}^2$ as a function of $K_1$ and $K_2$: the districts with less than 30\% of the population living in poverty.}\label{tabl3}%
	\resizebox{14cm}{!}{			
		\begin{tabular}{@{}lllllllllll@{}}
			\hline
			$K_1$ & $K_2=1$  & $K_2=2$ & $K_2=3$ & $K_2=4$  & $K_2=5$ & $K_2=6$ & $K_2=7$ & $K_2=8$  & $K_2=9$ & $K_2=10$ \\
			\hline
			1    & 0.0047   & --- & --- & --- & --- & --- & ---   & ---  &--- & --- \\
			2    & 0.0052   & \textbf{0.0041} & --- & --- & --- & --- & ---   & ---  &--- & --- \\
			3    & 0.0054   & \textbf{0.0039}  & 0.0045 & --- & --- & --- & ---   & ---  &--- & --- \\
			4    & 0.0052   & 0.0048 & 0.0045 & 0.0045 & --- & --- & ---   & ---  &--- & --- \\
			5    & 0.0048   & 0.0044 &  0.0043 & 0.0044 & 0.0048 &  --- & ---   & ---  &--- &---\\
			6    & 0.0044   & 0.0047  & 0.0047  & 0.0048 & 0.0050 & 0.0051 & ---   & ---  &--- &---\\
			7    & 0.0049   & 0.0045  & 0.0046 & 0.0049 & 0.0051 & 0.0052 & 0.0051 & ---  & --- &---\\
			8    & 0.0051 & 0.0043 & 0.0047 & 0.0051 & 0.0051 & 0.0051 & 0.0051 & 0.0051 & --- &---\\
			9    & 0.0047 & 0.0042 & 0.0046 & 0.0048 & 0.0050 & 0.0051 & 0.0050 & 0.0051 & 0.0052 & --- \\
			10    & 0.0053 & 0.0043 & 0.0047 & 0.0050 & 0.0051 & 0.0050 & 0.0050 & 0.0051  & 0.0051 & 0.0052 \\
			\hline
	\end{tabular} }
\end{table}

\begin{table}[h!]
	\centering
	\caption{The values of $\hat\sigma_{\epsilon}^2$ as a function of $K_1$ and $K_2$: the districts where 30\% - 55\% of the population live in poverty.}\label{tabl4}%
	\resizebox{14cm}{!}{
		\begin{tabular}{@{}lllllllllll@{}}
			\hline
			$K_1$ & $K_2=1$  & $K_2=2$ & $K_2=3$ & $K_2=4$  & $K_2=5$ & $K_2=6$ & $K_2=7$ & $K_2=8$  & $K_2=9$ & $K_2=10$ \\
			\hline
			1    & 0.0033   & --- & --- & --- & --- & --- & ---   & ---  &--- & --- \\
			2    & 0.0037 & \textbf{0.0027} & --- & --- & --- & --- & ---   & ---  &--- & --- \\
			3    & 0.0022 & \textbf{0.0021}  & 0.0035 & --- & --- & --- & ---   & ---  &--- & --- \\
			4    & 0.0026 & 0.0032 & 0.0039 & 0.0036 & --- & --- & ---   & ---  &--- & --- \\
			5    & 0.0039 & 0.0036 &  0.0038 & 0.0039 & 0.0041 &  --- & ---   & ---  &--- &---\\
			6    & 0.0049 & 0.0043  & 0.0044  & 0.0043 & 0.0045 & 0.0045 & ---   & ---  &--- &---\\
			7    & 0.0036 & 0.0038  & 0.0042 & 0.0042 & 0.0047 & 0.0048 & 0.0049 & ---  & --- &---\\
			8    & 0.0047 & 0.0037 & 0.0040 & 0.0044 & 0.0045 & 0.0047 & 0.0049 & 0.0049 & --- &---\\
			9    & 0.0048 & 0.0040 & 0.0044 & 0.0045 & 0.0046 & 0.0048 & 0.0048 & 0.0048 & 0.0049 & --- \\
			10    & 0.0042 & 0.0039 & 0.0046 & 0.0047 & 0.0047 & 0.0049 & 0.0049 & 0.0049  & 0.0050 & 0.0051 \\
			\hline
	\end{tabular} }
\end{table}

\begin{table}[h!]
	\centering
	\caption{The values of $\hat\sigma_{\epsilon}^2$ as a function of $K_1$ and $K_2$: the districts with more than 55\% of the population living in poverty.}\label{tabl5}%
	\resizebox{14cm}{!}{
		\begin{tabular}{@{}lllllllllll@{}}
			\hline
			$K_1$ & $K_2=1$  & $K_2=2$ & $K_2=3$ & $K_2=4$  & $K_2=5$ & $K_2=6$ & $K_2=7$ & $K_2=8$  & $K_2=9$ & $K_2=10$ \\
			\hline
			1    & 0.0078   & --- & --- & --- & --- & --- & ---   & ---  &--- & --- \\
			2    & 0.0079   & 0.0077 & --- & --- & --- & --- & ---   & ---  &--- & --- \\
			3    & 0.0077   & 0.0082  & 0.0076 & --- & --- & --- & ---   & ---  &--- & --- \\
			4    & 0.0083   & 0.0082 & 0.0075 & 0.0075 & --- & --- & ---   & ---  &--- & --- \\
			5    & 0.0082   & 0.0082 & 0.0066 & 0.0053 & 0.0046 &  --- & ---   & ---  &--- &---\\
			6    & 0.0081   & 0.0065 & 0.0046  & 0.0048 & 0.0045 & 0.0043 & ---   & ---  &--- &---\\
			7    & 0.0077   & 0.0046  & \textbf{0.0040} & 0.0045 & 0.0041 & 0.0042 & \textbf{0.0042} & ---  & --- &---\\
			8    & 0.0078 & 0.0063 & 0.0046 & 0.0048 & 0.0042 & 0.0043 & 0.0045 & 0.0045 & --- &---\\
			9    & 0.0082 & 0.0043 & 0.0044 & 0.0048 & 0.0046 & 0.0046 & 0.0046 & 0.0046 & 0.0048 & --- \\
			10    & 0.0082 & 0.0051 & 0.0044 & 0.0045 & 0.0045 & 0.0046 & 0.0046 & 0.0047  & 0.0048 & 0.0049 \\
			\hline
	\end{tabular} }
\end{table}
\medskip
As we have already mentioned, for the purpose of selecting $K_2$ districts, out of the $K_1$ previously selected districts, the variable ``Altitude'' is not the only variable that can be utilized in order to establish the degree of similarity between the districts. In the following table the results obtained in the case of utilizing the variables ``Poverty" and ``Extreme Poverty" are summarized. It can be concluded from Table \ref{tabl6} that the use of the variable ``Extreme Poverty" had some beneficial effect in the case of the first and the third group, while in the second group we would recommend to use the variable  ``Altitude". Notably, the variable ``Extreme Poverty" was not significant in the models presented in Table \ref{tabl2}. In summary, it can be inferred that choosing $K_2$ districts in the second step  using an additional variable to measure similarity between the $K_1$ previously selected districts, can potentially produce more powerful predictors (see Section 4.6).  

\begin{table}[h]
	\begin{center}
		\begin{minipage}{\textwidth}
			\caption{The optimal values of $\hat\sigma_{\epsilon}^2$ and the corresponding values of $K_1$ and $K_2$ for the variables ``Altitude", ``Poverty" and ``Extreme Poverty"}\label{tabl6}
				\begin{tabular*}{\textwidth}{@{\extracolsep{\fill}}lcccccc@{\extracolsep{\fill}}}
					\hline
					& \multicolumn{2}{@{}c@{}}{Less than 30\%} & \multicolumn{2}{@{}c@{}}{30\%-55\%} & \multicolumn{2}{@{}c@{}}{More than 55\%} \\
					Variable & Opt. $K_1,K_2$ & $\hat\sigma_{\epsilon}^2$ & Opt. $K_1,K_2$ & $\hat\sigma_{\epsilon}^2$ & Opt. $K_1,K_2$ & $\hat\sigma_{\epsilon}^2$\\
					\hline
					Altitude  & $(3,2)$ & 0.0039 & $(3,2)$ & 0.0021 & $(7,3)$ & 0.0040 \\
					Poverty  & $(3,2)$ & 0.0045  & $(2,1)$ & 0.0029 & $(7,2)$ & 0.0034 \\
					Ex. Poverty  & $(3,2)$ & 0.0033 & $(2,1)$ & 0.0027 & $(7,3)$ & 0.0032\\
					---  & $(2,2)$ & 0.0041 & $(2,2)$ & 0.0027 & $(7,7)$ & 0.0041\\
					\hline
				\end{tabular*}
		\end{minipage}
	\end{center}
\end{table}

\subsection{Spatial Fay-Herriot Model}
In what follows we fit the Spatial Fay-Herriot model for the following two scenarios.

\begin{enumerate}
	\item The neighbors are chosen using only the first step (the $K_1$ nearest neighbors), where $K_1=2$ for the districts with poverty level of less than 30\%, and for the districts with poverty level between 30\% and 55\%, and $K_1=7$ for the districts with poverty level of more than 55\%.  
	\item The neighbors are chosen using both steps, where in the second step we use the variable ``Extreme Poverty" for the districts with poverty level  of less than 30\% ($K_1=3, K_2=2$), and for the districts with poverty level of more than 55\% ($K_1=7, K_2=3$); for the districts with poverty level between 30\% and 55\%, the variable ``Altitude" was utilized with $K_1=3$ and $K_2=2$.   
\end{enumerate}

\noindent Tables \ref{tabl7} and \ref{tabl8} display the estimators for the coefficients $\beta$ and $\rho$ obtained by fitting the spatial Fay-Herriot model under the first and the second scenarios. The results illustrate that the spatial correlations are substentially high, especially for the poorer districts, being higher under the second scenario as opposed to the first scenario. This suggests that ignoring the spatial correlation structure between the districts may increase the potential for greater MSE. We can also conclude that the estimators for the coefficients $\beta$ are very similar under both scenarios. In comparing the results of this analysis with those presented in Table \ref{tabl2}, there is no drastic difference in the estimators. 

\begin{table}[h!]
	\caption{Estimators of the coefficients $\beta$ and $\rho$ of the spatial Fay-Herriot model under the first scenario}\label{tabl7}
		\begin{tabular*}{\textwidth}{@{\extracolsep{\fill}}lcccccc@{\extracolsep{\fill}}}
			\hline%
			& \multicolumn{2}{@{}c@{}}{Less than 30\%} & \multicolumn{2}{@{}c@{}}{30\%-55\%} & \multicolumn{2}{@{}c@{}}{More than 55\%} \\
			& Estimator & p.value & Estimator & p.value & Estimator & p.value \\
			\hline
			Water  & --- & --- & -0.12640 & 0.0041 & -0.07681 & 0.0863 \\
			Water-days  & 0.10153 & 0.0281  & --- & --- & --- & --- \\
			Floor  & -0.09349 & 0.041 & --- & --- & --- & ---\\
			Refrig.  & --- & --- & -0.34632 & $<0.0001$ & -0.22646 & 0.0002\\
			Internet  & -0.30843 & $<0.0001$ & --- & --- & --- & ---\\
			Spanish  & -0.24583 & $<0.0001$ & -0.17426 & $<0.0001$ & -0.20250 & $<0.0001$\\
			SIS  & --- & --- & -0.14858 & 0.0494 & --- & ---\\
			Uninsur. & --- & ---  & ---  & --- & 0.21169 & 0.0926\\
			Altitude & 0.00002 & 0.0122  & --- & --- & 0.00001 & 0.6277\\
			Rural & --- & --- & -0.07971 & 0.0208 & --- & ---\\
			$\rho$ & 0.4495 & $<0.0001$ & 0.6548 & $<0.0001$ & 0.8062 & $<0.0001$ \\
			\hline
		\end{tabular*}
\end{table}

\begin{table}[h!]
	\caption{Estimators of the coefficients $\beta$ and $\rho$ of the spatial Fay-Herriot model under the second scenario}\label{tabl8}
		\begin{tabular*}{\textwidth}{@{\extracolsep{\fill}}lcccccc@{\extracolsep{\fill}}}
			\hline%
			& \multicolumn{2}{@{}c@{}}{Less than 30\%} & \multicolumn{2}{@{}c@{}}{30\%-55\%} & \multicolumn{2}{@{}c@{}}{More than 55\%} \\
			& Estimator & p.value & Estimator & p.value & Estimator & p.value \\
			\hline
			Water  & --- & --- & -0.10861 & 0.0144 & -0.08323 & 0.0863 \\
			Water-days  & 0.08005 & 0.0735  & --- & --- & --- & --- \\
			Floor  & -0.09976 & 0.0286 & --- & --- & --- & ---\\
			Refrig.  & --- & --- & -0.34411 & $<0.0001$ & -0.20772 & 0.0004\\
			Internet  & -0.29959 & $<0.0001$ & --- & --- & --- & ---\\
			Spanish  & -0.24556 & $<0.0001$ & -0.17884 & $<0.0001$ & -0.19998 & $<0.0001$\\
			SIS  & --- & --- & -0.13017 & 0.0888 & --- & ---\\
			Uninsur. & --- & ---  & ---  & --- & 0.20459 & 0.0940\\
			Altitude & 0.00002 & 0.0253  & --- & --- & 0.00001 & 0.4073\\
			Rural & --- & --- & -0.07049 & 0.0484 & --- & ---\\
			$\rho$ & 0.4984 & $<0.0001$ & 0.7275 & $<0.0001$ & 0.8144 & $<0.0001$ \\
			\hline
		\end{tabular*}
\end{table}
\medskip

In the following section we compare the EBLUP and the SEBLUP as well as their corresponding MSEs.

\subsection{EBLUP, SEBLUP and MSE}

First, we compare the predictions EBLUP and SEBLUP for the prevalence of anemia rates among children under five years, with the corresponding direct estimates. In the following tables, SEBLUP1 and SEBLUP2 refer to the predictors SEBLUP obtained under the first and the second scenarios defined above. For the purpose of these comparisons the following three groups of districts are used: the districts that only ahve 5 observations (a total of 15 districts), the districts with 15 observations (a total of 16 districts) and the districts with 40-49 observations (a total of 24 districts). 
\medskip

\begin{figure}[h]
	\centering
	\includegraphics[width=14cm]{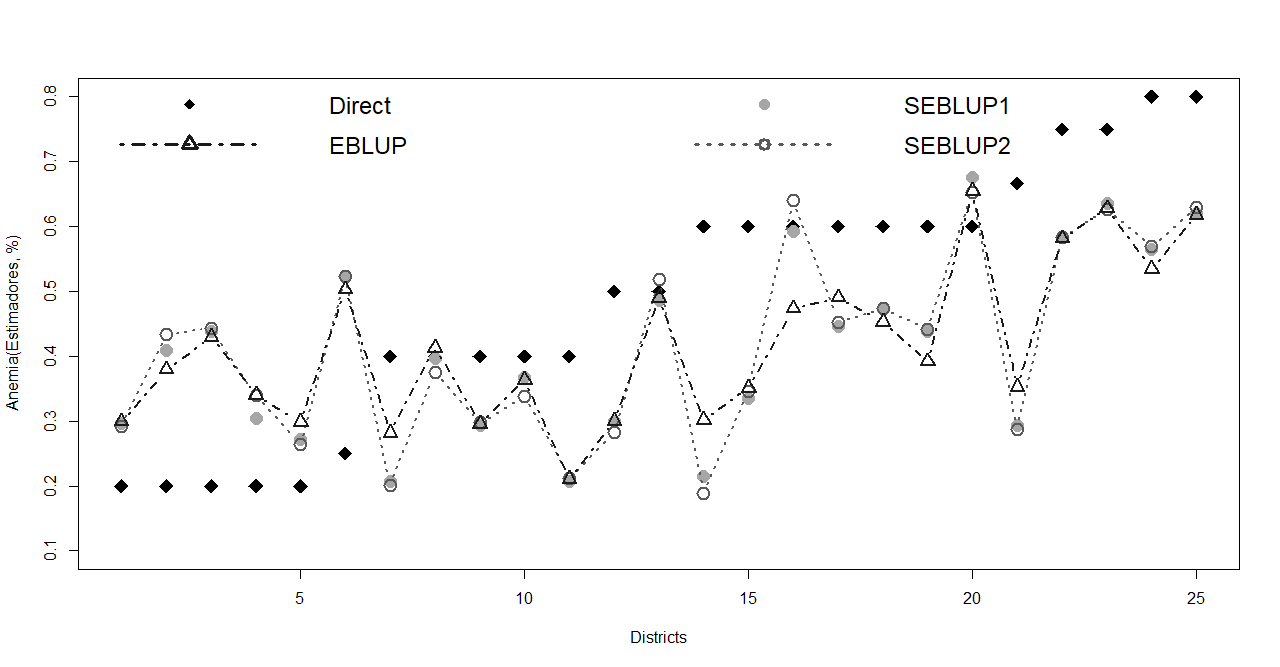} 
	\caption{Estimates for the prevalence of anemia rates in the districts with 5 observations}\label{est5}	
\end{figure}

\begin{figure}[t!]
	\centering
	\includegraphics[width=14cm]{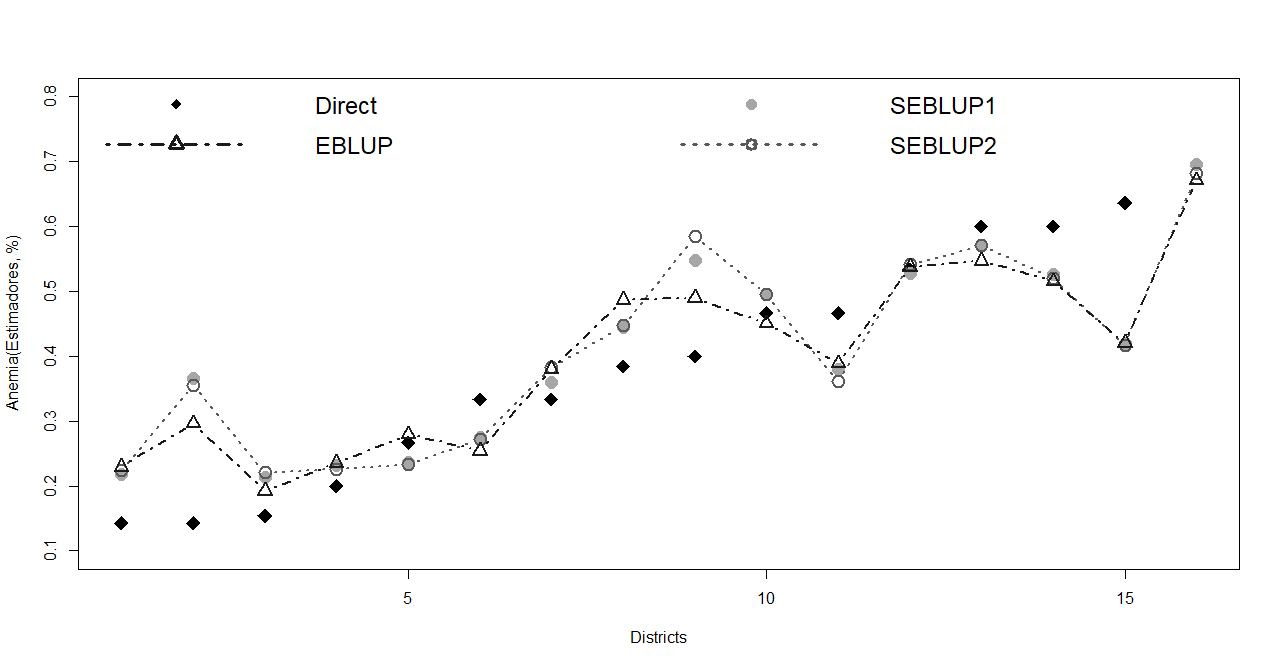} 
	\caption{Estimates for the prevalence of anemia rates in the districts with 15 observations}\label{est15}
\end{figure}

\begin{figure}[h!]
	\centering
	\includegraphics[width=14cm]{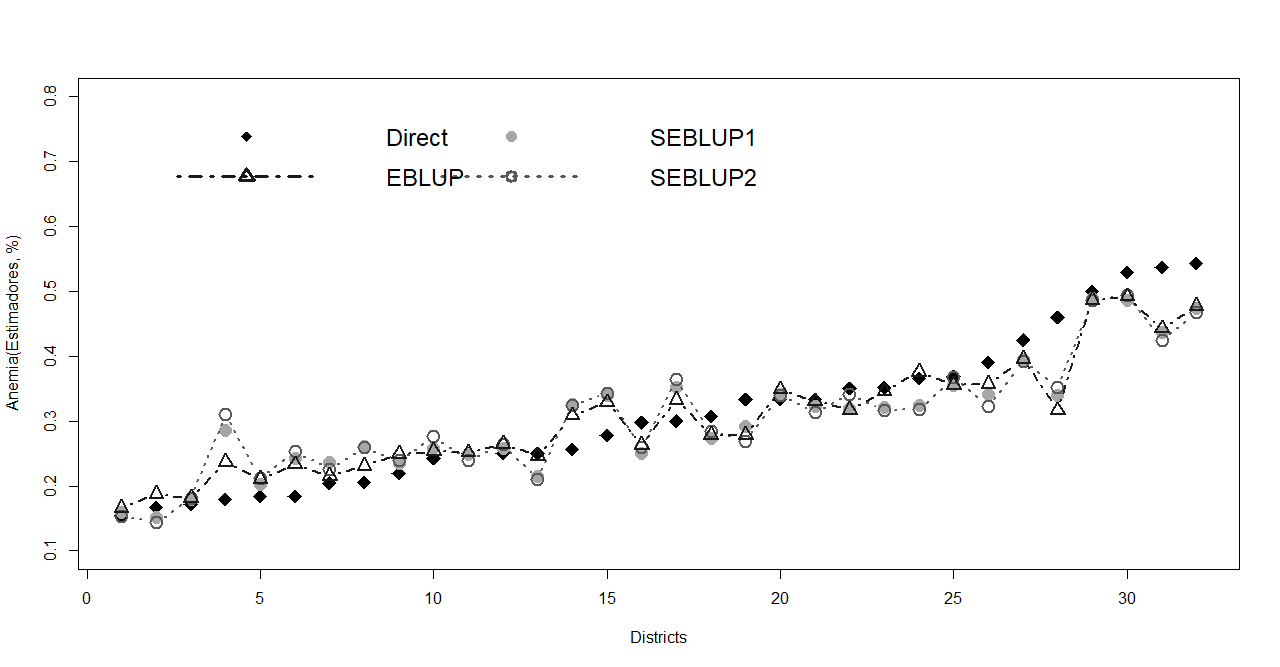} 
	\caption{Estimates for the prevalence of anemia rates in the districts with 40-49 observations}\label{est40}
\end{figure}

As expected, the results presented in Figures \ref{est5}-\ref{est40} illustrate that the differences between SEBLUP1, SEBLUP2, EBLUP and the corresponding direct estimate decreases as the sample size increases. Interestingly, the  discrepancies between SEBLUP1 and SEBLUP2 are generally minor: the mean absolute differences between EBLUP1 and EBLUP2 is 0.014 in the first case, 0.010 in the second case and 0.009 in the third case. The corresponding relative differences amount to 3.6\%, 2.4\% and 3.1\%,  respectively.  

\break

\break

\medskip

\begin{figure}[h!]
	\centering{
		\includegraphics[width=13.9cm]{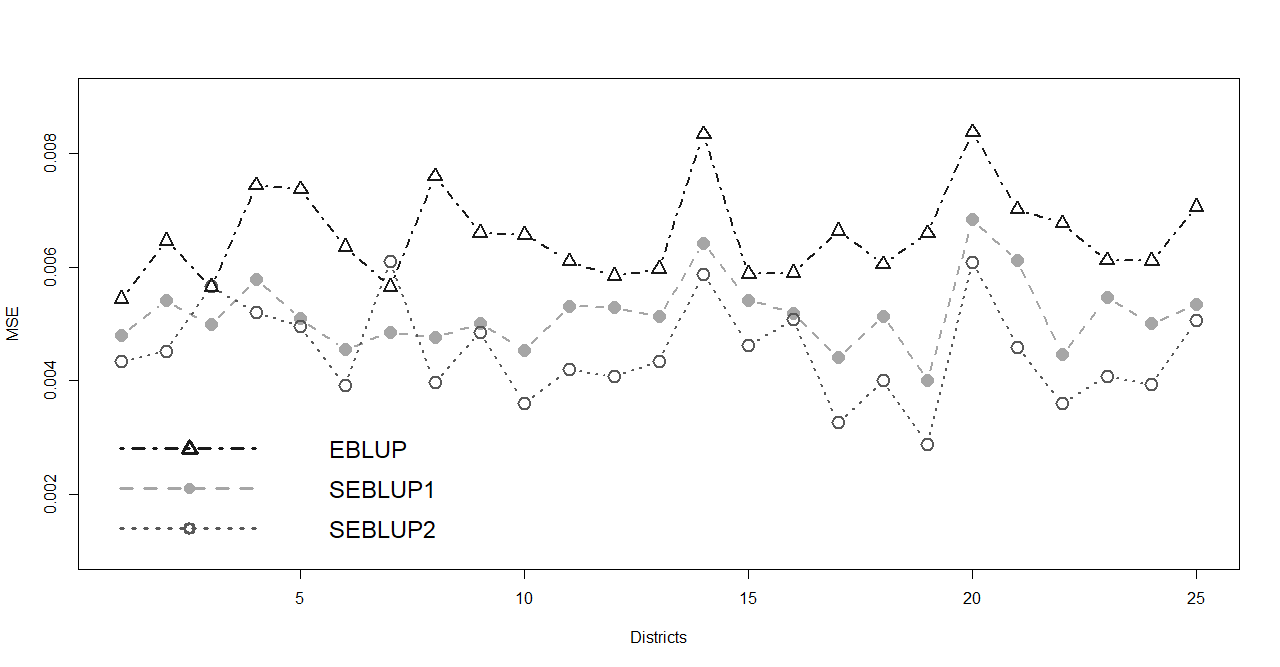} 
		\caption{Mean Square Errors of the estimates for the prevalence of anemia rates in the districts with 5 observations}\label{mse5}
		
	}
\end{figure}

\begin{figure}[h!]
	\centering{
		\includegraphics[width=13.9cm]{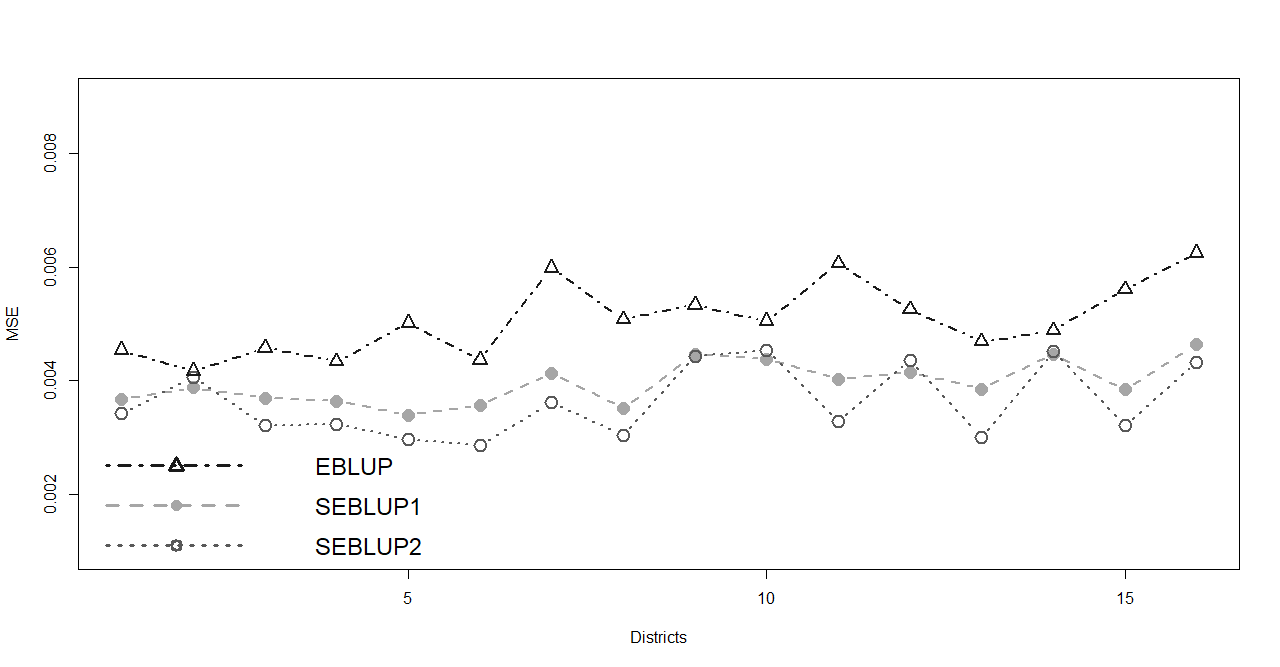} 
		\caption{Mean Square Errors of the estimates for the prevalence of anemia rates in the districts with 15 observations}\label{mse15}
	}
\end{figure}

\begin{figure}[h!]
	\centering{
		\includegraphics[width=13.9cm]{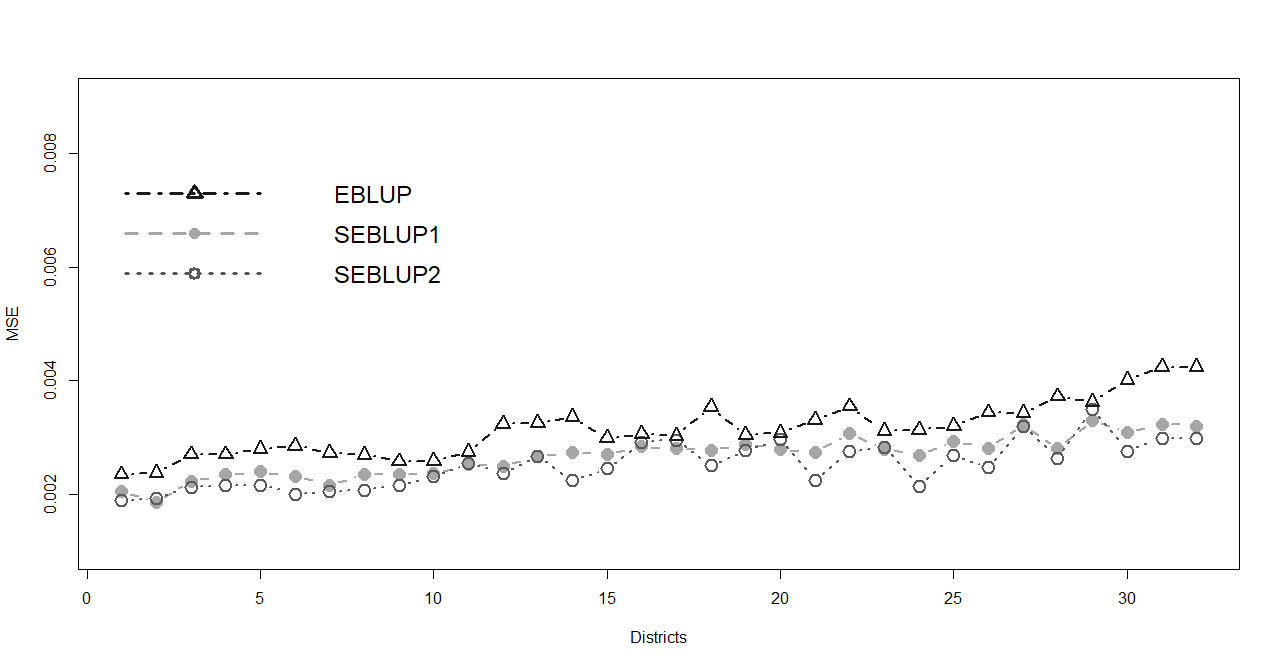} 
		\caption{Mean Square Errors of the estimates for the prevalence of anemia rates in the districts with 40-49 observations}\label{mse40}
	}
\end{figure}

Next, we present the MSEs of the discussed predictors. Figures \ref{mse5}-\ref{mse40} display the MSEs obtained by application of the parametric bootstrap. The MSEs derived from application of the non-parametric bootstrap are somewhat larger, however, the conclusions reached are very similar to those reported below. The results indicate very clearly that in our case application of the spatial Fay-Herriot model yields better MSEs than the basic Fay-Herriot model. The results also provide evidence that except for several districts, the MSEs of SEBLUP2 have had better performance than SEBLUP1 and EBLUP,  especially if the sample size is small. Specifically, the relative difference in MSE between SEBLUP1 and SEBLUP2 are 12.9\%, 8.8\% and 6.0\% in the first, second and third case, respectively. 

\newpage

Finally, we compute the coefficients of variation (CV) for all predictors discussed above. As one can observe from Table \ref{tabl9}, the direct estimator has very large CV in the districts where the sample size is smaller than 50. If the sample size is larger than 50, only for 61 districts (out of 104 districts) the CV of the direct estimator is smaller than 20\%. Comparing this result to the corresponding numbers for EBLUP (87 districts), SEBLUP1 (92 districts) and SEBLUP2 (92 districts), we can conclude that for large samples employing the basic Fay-Herriot as well as the spatial Fay-Herriot considerably improve the precision of the predictors, where the SEBLUP1 and SEBLUP2 slightly outperform the EBLUP. For smaller sample sizes we can observe a similar pattern; the difference is that in these cases the performance of SEBLUP1 and SEBLUP2 is much better than that of the EBLUP, especially if the sample size is less than 7 or between 7 and 10. Moreover, the performance of SEBLUP2 is evidently better for all sample sizes.  

\begin{table}[h!]
	\begin{center}
		\caption{Distribution of coefficient of variation for Direct estimate, EBLUP, SEBLUP1 and SEBLUP2 by sample size}\label{tabl9}%
			\begin{tabular}{@{}lllllll@{}}
				\hline
				Sample Size & Predictor &  $<10\%$  & 10-20\% & 20-30\% & $>30\%$ & Total\\
				\hline
				Less than 7    & Direct   & 0 & 1 & 4 & 120 & 125\\
				Less than 7    & EBLUP   & 0 & 57 & 56 & 12 & 125\\
				Less than 7    & SEBLUP1   & 0 & 78 & 40 & 7 & 125\\
				Less than 7    & SEBLUP2   & 0 & 82 & 40 & 3 & 125\\
				7-10    & Direct   & 0 & 11 & 20 & 164 & 195\\
				7-10    & EBLUP   & 0 & 76 & 89 & 30 & 195\\
				7-10    & SEBLUP1   & 1 & 110 & 61 & 23 & 195\\
				7-10    & SEBLUP2   & 1 & 120 & 57 & 17 & 195\\
				11-20    & Direct   & 0 & 16 & 50 & 161 & 227\\
				11-20    & EBLUP   & 1 & 106 & 90 & 30 & 227\\
				11-20    & SEBLUP1 & 2 & 128 & 79 & 18 & 227\\
				11-20    & SEBLUP2   & 3 & 145 & 63 & 16 & 227\\
				21-50    & Direct   & 0 & 27 & 65 & 69 & 161\\
				21-50    & EBLUP   & 0 & 84 & 65 & 12 & 161\\
				21-50    & SEBLUP1   & 2 & 104 & 46 & 9 & 161\\
				21-50    & SEBLUP2   & 6 & 113 & 34 & 8 & 161\\
				More than 50 & Direct   & 10 & 51 & 39 & 4 & 104\\
				More than 50 & EBLUP   & 10 & 77 & 15 & 1 & 104\\
				More than 50  & SEBLUP1   & 11 & 81 & 11 & 1 & 104\\
				More than 50 & SEBLUP2   & 13 & 79 & 12 & 0 & 104\\
				All Districts & Direct   & 10 & 106 & 178 & 518 & 812\\
				All Districts & EBLUP   & 11 & 400 & 315 & 86 & 812\\
				All Districts  & SEBLUP1   & 16 & 501 & 237 & 58 & 812\\
				All Districts & SEBLUP2   & 23 & 539 & 206 & 44 & 812\\
				\hline
			\end{tabular}
	\end{center}
\end{table}

\section{Conclusion}\label{sec13}
From the results obtained in Section 4 we conclude that utilizing the basic Fay-Herriot model have considerably removed the MSE (and therefore, the CVs) of the predictors as opposed to the direct estimates. However,  the obtained CVs in most of the districts are still substentially large. If the spatial Fay-Herriot model is applied, an additional reduction in MSEs is attained. This is due to incorporating information about the spatial structure of the data, which is ignored by the basic model. The reduction in MSE is more substential if we select the neighbors using the two-step procedure which allows to employ additional information about the districts (see Section 4.2). Regarding the question about reliability of the EBLUP and SEBLUP, the magnitudes of the corresponding CVs indicate that in the first case the percentage of unreliable estimates (the estimates with the CV larger than 20\%) is considerably large, especially if the sample is small. Specifically, if the sample size is smaller than 7, the percentage of unreliable estimates is 54\%. For larger sample sizes we observe a very modest reduction (48\% if the sample size is between 21 and 51). If the sample size is larger than 50, the percentage of unreliable estimators reduces to 15\%. In the case of the SEBLUP the corresponding percentages are as follows: 34\% if the sample size is smaller than 7, 26\% if the sample size is between 21 and 50 and 12\% if the sample size is larger than 50. However, it should be noticed that the percentage of the estimates whose CV is larger than 30\% is relatively small: in the case of the EBLUP it oscillates between 7 and 15\% (for SEBLUP the range is between 2 and 9\%) if the sample size is smaller than 50. If the sample size is larger than 50, the CV of only 1 predictor EBLUP (out of 104) is larger than 30\%. In the case of the SEBLUP, the CVs of all predictors is smaller than 30\%. Apart from comparing the performance of the basic and the spatial Fay-Herriot models we explore the sensitivity of the choice of the neighbors to the resulting inference. It follows from the results that the conclusions drawn can depend significantly on the definition of the neighbors. We recommend that, in practice, one chooses the definition that acheive the smallest variance, $\hat\sigma_{\epsilon}^2$. There is no theoretical basis for this choice, however it may be advantageous from the perspective of reduction of the MSEs of the predictors. In this paper we do not discuss the problem of prediction in the nonsampled district. This can be a topic for future research.

\subsection*{Acknowledgements.} 
The views presented in this work are those of the authors and do not represent the official position of the institutions that the authors are or were affiliated with. This research is supported by a grant from the Unidad de Investigaci\'on, FIEECS-UNI.

\end{document}